\documentclass[preprint,aps,showpacs,preprintnumbers,amsmath,amssymb]{revtex4}  

\usepackage{graphicx}
\usepackage{dcolumn}
\usepackage{bm}
\usepackage{color}

\begin{document}

\title{The effect of inelasticity on the phase transitions of a thin vibrated granular layer}

\author{Francisco Vega Reyes}
  \affiliation{Departamento de F\'isica, Universidad de Extremadura,
Avda. Elvas s/n, 06071 Badajoz, Spain}
\author{Jeffrey S. Urbach}
\email{urbach@physics.georgetown.edu}
\affiliation{Department of Physics, Georgetown University, Washington DC 20057, USA.}

\date{November 4, 2008}

\begin{abstract}
We describe an experimental and computational investigation of the ordered and disordered phases of a vibrating thin, dense granular layer composed of identical metal spheres. We compare the results from spheres with different amounts of inelasticity and show that inelasticity has a strong effect on the phase diagram. We also report the melting of an ordered phase to a homogeneous disordered liquid phase at high vibration amplitude or at large inelasticities. Our results show that dissipation has a strong effect on ordering and that in this system ordered phases are absent entirely in highly inelastic materials.
\end{abstract}

\maketitle

Ordered phases observed in non-cohesive granular media demonstrate both profound similarities to and differences from those observed in elastic systems, where the results of equilibrium statistical mechanics apply \cite{reis06,SprBook,JCP,clerc08,watanabe08}. Controlled experiments on simple model systems are necessary to develop and test extensions of  statistical mechanics to non-equilibrium systems \cite{aranson06}. In previous research \cite{PRLord,PREcub,JCP,PRLmelt}, we have reported the existence of ordered phases in vibrated dense thin layers of identical metal spheres. For monolayers at high densities, hexagonal ordering is observed \cite{PRLord}, and recent results from our group \cite{PRLmelt} and, in a somewhat different setup, Reis et al. \cite{reis06}, show that the crystallization of the ordered phase can be directly mapped onto the analogous 2D equilibrium system.  In the presence of a confining lid, we have reported a complex phase diagram that is closely related to that observed in similarly confined equilibrium colloidal systems, including two-layer crystals with square or hexagonal symmetry \cite{JCP,PREcub}. In a recent study Clerc  et al. \cite{clerc08} extended this work to quasi-one-dimensional systems, and showed that the transition was mediated by traveling waves and was triggered by negative compressibility.   

In this paper we present the experimental phase diagram for the confined granular layer in more detail than reported previously, for both stainless steel and brass spheres.   We report the existence of a melting transition of the ordered phase as the vibration amplitude is increased, and show that similar behavior is observed in computer simulations of inelastic spheres. The melting of the crystalline phase occurs significantly earlier in the system of brass spheres, and computer simulations show that the ordered phase disappears entirely in the presence of high inelasticity.  This result demonstrates that dissipation can dominate over geometric packing effects for sufficiently inelastic spheres.

\begin{figure}[h]
  \includegraphics[height=1.5cm]{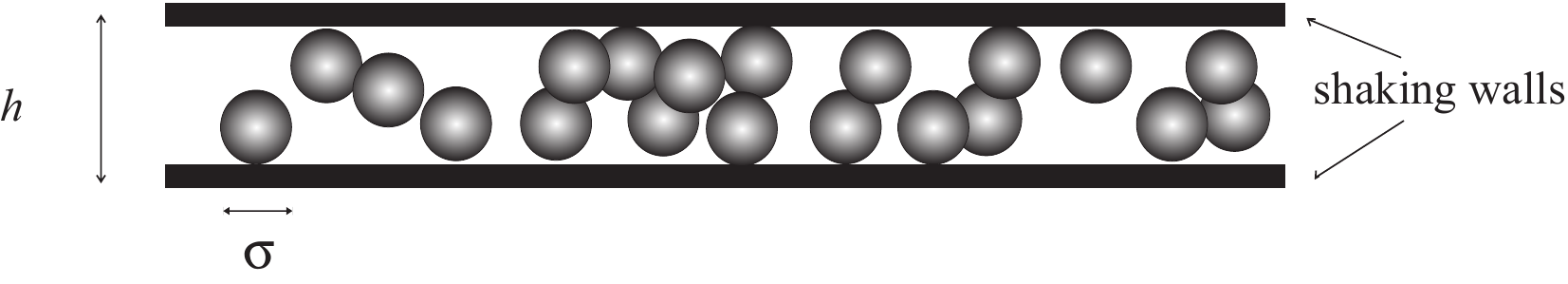}
\caption{Schematic side view of the experimental setup, consisting of two horizontal planes, separated by a gap of $h=1.75~\sigma$. Both walls vibrate together sinusoidally in the vertical direction.} \label{setup}
\end{figure}

\begin{figure*}
\includegraphics[width=15cm]{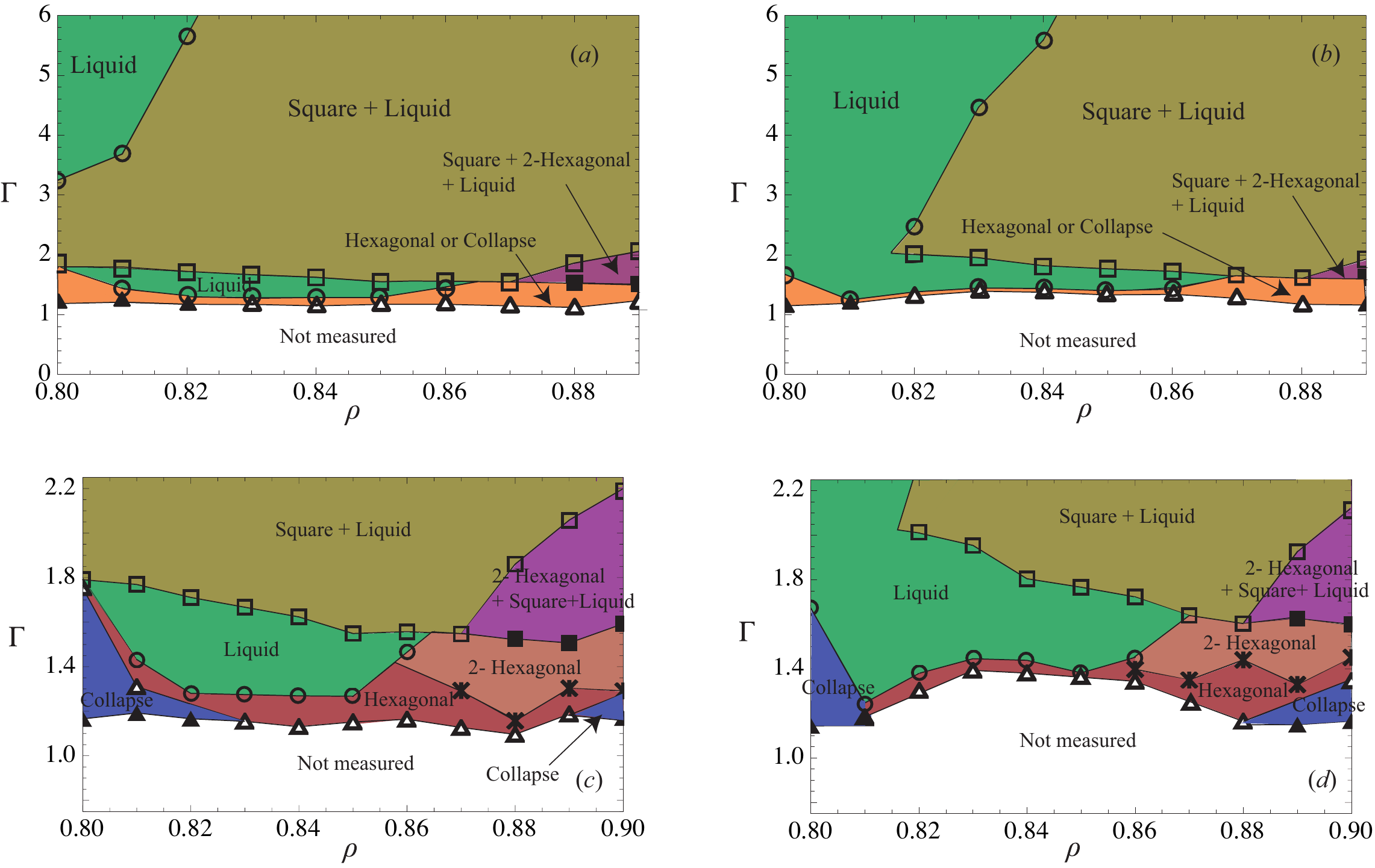}
\caption{Experimental phase map for steel (\textit{a}) and brass (\textit{b}) spheres. The critical reduced acceleration $\Gamma$, for which a phase appears is plotted against the 2D density. We find the following phases: Hexagonal-collapse ($\blacktriangle$), hexagonal-liquid ($\triangle$), double-hexagonal-liquid ($\ast$), liquid ($\bigcirc$), square-liquid-hexagonal ($\blacksquare$), square-liquid ($\square$). In the lower panels the phase maps for steel and brass respectively are expanded in the $\Gamma\sim 1-2$ region. } \label{expfase}
\end{figure*}

The experimental system consists of nearly identical metallic spheres confined in a gap which lies horizontally between a circular plate and a lid (Fig. \ref{setup}).  In this paper we report results with stainless steel and brass spheres with diameter $\sigma=1.5875\pm0.0032$ mm ($1/16''$). The plate diameter is $168$ mm, or equivalently, $d_p=112~\sigma$, and the gap spacing is $h=2.78$ mm $=1.75~\sigma$ for all experiments. We characterize the system's density with the 2D density $\rho$, defined as $\rho=N/N_{max}$, where $N_{max}=11377$ is the maximum number of balls that can fit in a hexagonally packed monolayer of balls on the plate at rest. In the experiments reported here the frequency of the sinusoidal plate vibration, provided by an electromagnetic shaker, was fixed at $\nu=60$ Hz. In what follows we report the amplitude of the oscillation, $A$, as the reduced acceleration $\Gamma=A\omega^2/g$, where $\omega=2\pi \nu$ and $g$ is the acceleration due to gravity. The acceleration is measured by a fast response accelerometer mounted on the bottom plate and maintained at a constant value with a computer-based feedback loop. 

Under experimental conditions similar to our setup, it has been shown that the particle-particle collisions can be accurately described by a model characterized by three coefficients: $e$, which characterizes the incomplete restitution in the normal component of the relative particle velocities in the collision; $\beta_0$, which is the tangential coefficient of restitution for non-sliding collisions; and $\mu$, which is a frictional coefficient for sliding collisions \cite{LougeModel}. Brass and steel spheres differ primarily in their coefficients of normal restitution $e$: for steel, $e=0.95$; and for brass, $e=0.77$ \cite{Impact}. Therefore, brass balls lose approximately 4 times more energy per unit mass in particle-particle collisions, compared to steel balls, since the kinetic energy lost by the normal component of relative velocities is proportional to $1-e^2$ \cite{Goldhirsch} ($1-e^2=0.0975$ for steel whereas $1-e^2=0.4071$ for brass).
Thus, the system of brass spheres is significantly further from equilibrium than the comparable system of stainless steel spheres.

In order to map out an experimental phase diagram, we started from an acceleration close to 1 $g$ and increased the amplitude in small increments $\Delta\Gamma\sim 0.025~g$. After each increment, we waited until there was no discernible evolution of the system, and then waited several minutes longer to ensure that a steady state had been reached.  The phases present were then determined by visual inspection. This procedure was repeated for a range of densities, for both steel and brass spheres. The results are summarized in Fig. \ref{expfase}. Figs. \ref{expfase}a and b show the phase diagrams obtained for steel and brass, respectively, while panels (c) and (d) show expanded views of the region below $\Gamma=2$. 

At the lowest amplitudes, we observe a `collapse' of motionless, hexagonal close-packed spheres 
\cite {PRLord}. At slightly higher acceleration, the spheres order into a fluctuating hexagonally ordered singly layer \cite{PRLmelt}. As the acceleration is increased further, the hexagonal phase melts into a homogeneous liquid.  If we continue to increase input acceleration, small clusters, denser than the surrounding fluid, begin to appear, initially unstable in time. With further increase of input acceleration one of the clusters becomes more stable and nucleates a two-layer ordered phase with square symmetry \cite{PREcub}. These phase diagrams are consistent with previous reports, but are more comprehensive and represent the first direct comparison between different types of particles.  

This experimental phase diagram shows two notable new features. The first one is that for some densities the square phase melts at high vibration amplitude, and the second one is that for some parameter ranges the ordered square phase present in steel spheres is completely absent in brass.  (For instance the square phase forms for steel spheres in the range of density $\rho=0.80-0.815$, while it is not present for brass spheres for densities below $\rho=0.82$ for any vibration amplitude). While much of the phase behavior of this system can be understood by analogy with equilibrium colloidal systems \cite{PREcub,JCP}, the meting of the square phase due to increase of acceleration and/or inelasticity reported here are purely nonequilibrium effects. We have speculated that some of the behavior we have observed upon increasing acceleration may be due to layer compression \cite{PREcub,JCP}. It is not clear why this effect would be more pronounced in more inelastic material, except that it is clearly somehow associated with deviations from elastic behavior. Computer simulations by Clerc et al. \cite{clerc08} of a confined 2D (hard disk) system showed that increasing inelasticity reduced the liquid-solid coexistence region, consistent with the results reported here.

\begin{figure}
\includegraphics[height=4.5cm]{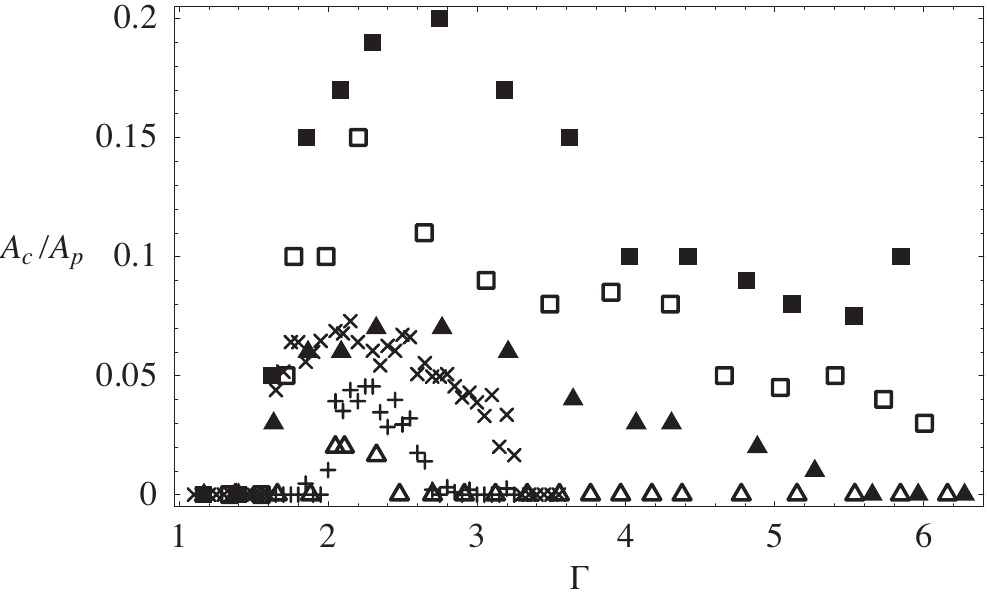}
\caption{The area of the square phase, $A_c$, divided by the plate area, $A_p$,  in experiments with densities $\rho=0.82$ (triangles) and $\rho=0.86$ (squares) for both brass (open symbols) and steel balls (solid symbols), and MD simulations ($\rho=0.90$ and $\gamma_n=200~s^{-1}$  ($\times$) and $\gamma_n=262.5~s^{-1}$ ($+$),  $N=2000$) as a function of input acceleration.}
\label{tam}
\end{figure}

To further investigate the dependence of the phase coexistence on acceleration and inelasticity, we have measured the area occupied by the square phase as a function of input acceleration, from its appearance up to its disappearance, for both brass and steel spheres. The results for two different densities are shown in Fig. \ref{tam}.  In all cases the freezing transition is abrupt, consistent
with previous observations \cite{PREcub}, while the disappearance is gradual. For both density values, the square phase maximum size is lower for the more inelastic brass spheres and it is observed over a smaller range of accelerations.  

In order to confirm that the inevitable imperfections in the experimental setup are not responsible for the features observed in the phase map and to more thoroughly study the effect of varying inelasticity, we have also analyzed a series of molecular dynamics simulations (MD), using the same procedure as in \cite{PREcub}. The collisions are characterized by three forces: two normal forces (one elastic restoring force, proportional to particle overlap, and one frictional dissipative force, proportional to relative normal velocity) and a tangential force (inelastic). We have used values of parameters in the simulation that mimic the behavior of stainless steel balls (see \cite{ChenMD} for details), and then vary the normal frictional force by varying the constant of proportionality, $\gamma_n$, to investigate the effect of increased inelasticity.  The interactions included in the model do not capture the full complexity of real inelastic collisions, but reproduce the dominant effects of vibration, collisions, and dissipation, and show all of the same qualitative features of the experiment.  

The simulations followed the same sequence as the laboratory experiments.  First, an initial state is prepared by placing spheres at random positions, with the constraint that particles do not overlap.   Particles are assigned random velocities chosen from a uniform distribution.   The simulation then runs at constant frequency and amplitude until a steady state is reached (as indicated by a constant granular temperature), which typically takes a few seconds of simulated time.    Then the acceleration was increased in small increments, $\Delta\Gamma=0.025~g$, keeping the frequency constant, with sufficient delay between each increase to ensure that a steady state was again reached.   As reported previously \cite{PREcub}, the simulations reproduce the ordered phases observed experimentally.  The crosses in Fig. \ref{tam}a show the size of the square phase as a function of acceleration for one value of the normal dissipative force parameter, $\gamma_n=200~s^{-1}$, and the plusses show the size for a larger value, $\gamma_n=262.5~s^{-1}$, for a simulation with $N=2000$ particles.  

\begin{figure}
\includegraphics[height=11.75cm]{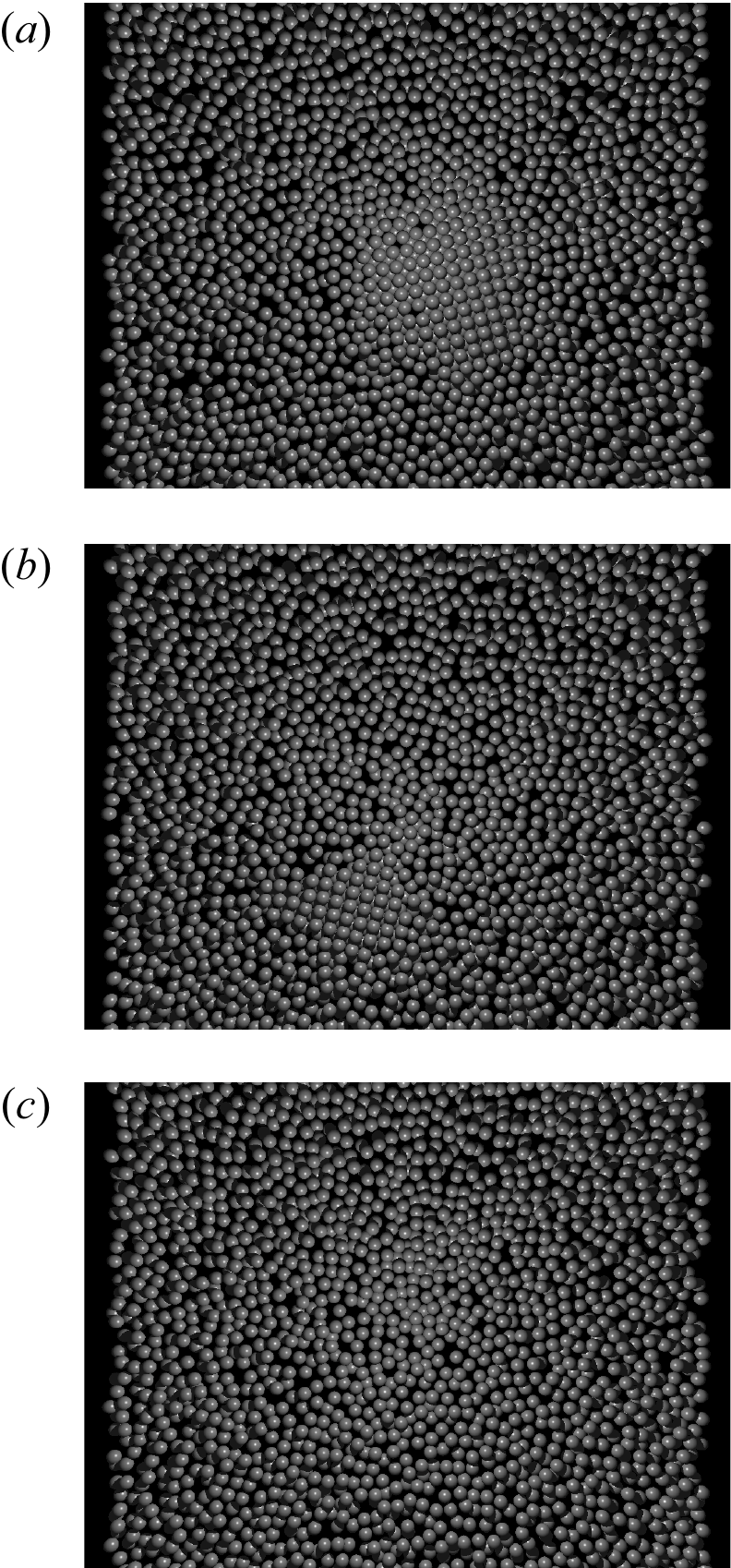}
\caption{Renderings of MD simulations snapshots, for different values of the vibration amplitude. In (\textit{a}), with $\Gamma=1.85$, there is a stable ordered two-layer square phase coexisting with the disordered phase. (\textit{b}) A snapshot $\sim 3~$s after a sudden acceleration increase to $\Gamma=3.5$, showing the shrinking unstable crystal.  (\textit{c}) The steady state for $\Gamma=3.5$, where the only present phase is a homogeneous liquid. ($\gamma_n=200$ s$^{-1}$, $\rho=0.895$, $N=2000$).}
\label{simfase}
\end{figure}

The simulation reproduces the behavior of the experiments:  the ordered phase disappears for large acceleration amplitudes, and the region of stability is smaller for more inelastic spheres.  An example of the evolution upon increasing acceleration is shown in Fig. \ref{simfase}.  Fig. \ref{simfase}a shows a rendering of the sphere positions for $\Gamma=1.85$, where the square phase is stable.  Upon sudden increase in acceleration amplitude to $\Gamma=3.5$, the crystal shrinks (Fig. \ref{simfase}b) and then disappears entirely, leaving only a disordered state (Fig. \ref{simfase}c).

\begin{figure}
\includegraphics[height=8.5cm]{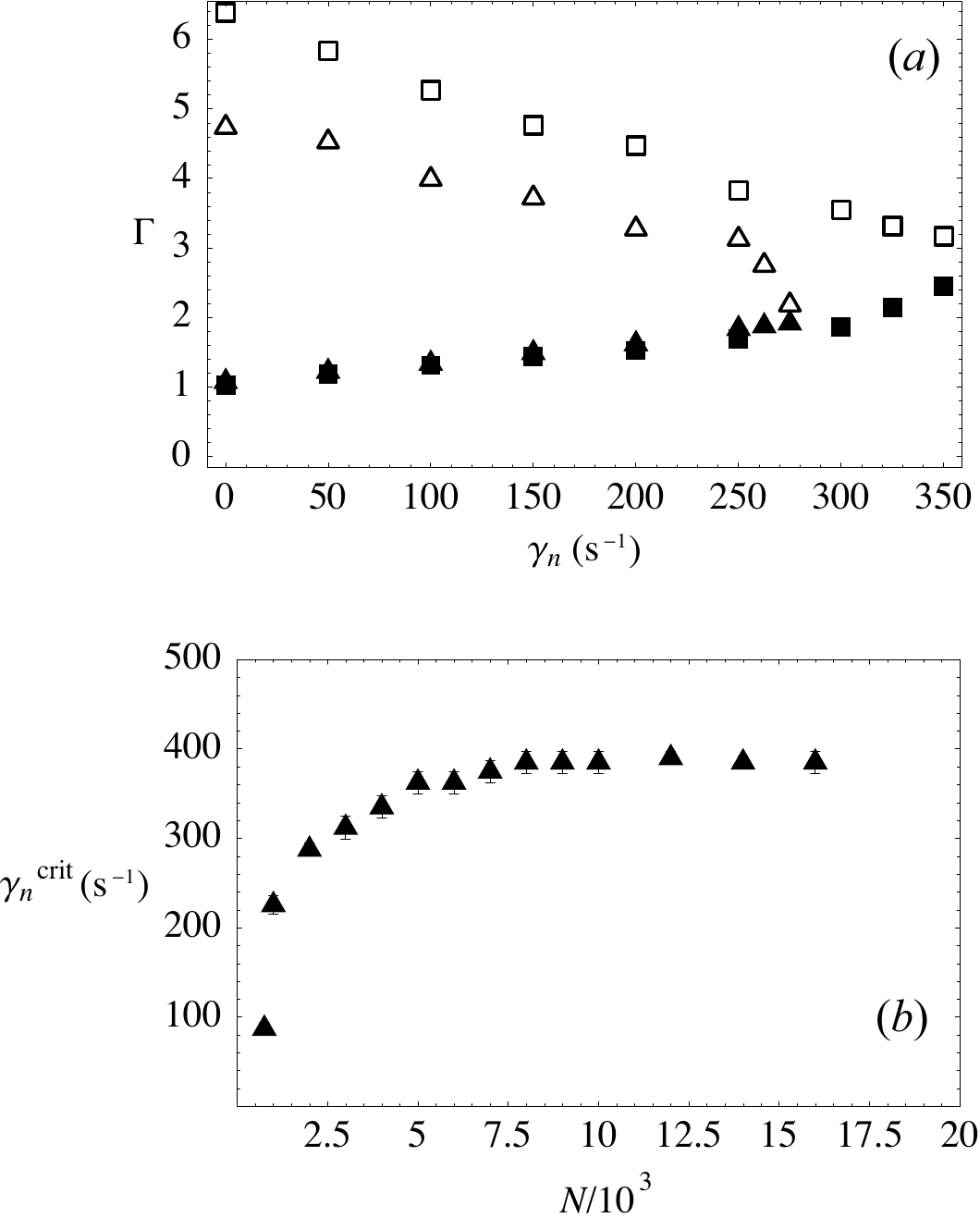}
\caption{(\textit{a})     Critical acceleration values from MD simulations for square phase nucleation (solid symbols) and melting (open symbols), for varying $\gamma_n$, for simulations with $N=2000$ (triangles)  and $N= 6000$ (squares)  (\textit{b}) Maximum value of $\gamma_n$ for existence of the square phase in MD simulations with different values $N$. ($\rho=0.895$, $h=1.75~\sigma$, $\nu=60$ Hz.)}
 \label{Gg}
\end{figure}

Figure \ref{Gg}a shows the critical value of the acceleration for the formation of the square phase at $\rho=0.895$ upon increasing acceleration (solid symbols) and for its melting (open symbols) as a function of inelasticity, for $N=2000$ (triangles) and $N=6000$ (squares).  For both values of $N$, the region of stability of the crystal decreases as the inelasticity increases, as does the maximum size of the crystal (data not shown).  Above a critical value  of $\gamma_n$ ($\gamma_n^{crit}$), the crystal does not form for any vibration amplitude.  Figure \ref{Gg}a also shows that the crystallization transition depends on the size of the simulated system. Finite-size effects are well known in constant volume simulations of equilibrium transitions \cite{wedekind06}, and arise because the formation of a dense phase exceeding the critical nucleation size results in a finite reduction in the density of the coexisting disordered phase.  The appearance of finite-size effects in the granular crystallization suggests that something analogous to a strong surface tension is present, despite the absence of attractive interactions.  The origin of this effect is unknown, but we note that the granular temperature in the crystal is higher near the edges than in the center \cite{PREcub}, and this `boundary layer' may contribute to an effective surface tension.  

In Fig. \ref{Gg}b we plot the value of $\gamma_n^{crit}$ as a function of system size.  The results clearly show that the finite-size effects become negligible for sufficiently large systems, and that there is a critical value of inelasticity above which the ordered phase is suppressed for any system size.  This surprising result 
provides a vivid demonstration that free-volume considerations that explain the transitions observed in confined equilibrium hard-sphere systems are not sufficient to describe the behavior of inelastic spheres, despite the strong similarities. We have observed previously that the granular temperature is significantly lower in the crystal than in the coexisting liquid \cite{PREcub}, and suggested that this difference may account for the presence of a coexistence region significantly larger than that found in equilibrium systems \cite{JCP}.  To investigate whether the absence of equipartition was also playing a role in the suppression of the crystalline phase, we measured the granular temperature in the ordered and liquid phases in the MD simulations reported here. The temperatures of the phases do change with the dissipation, but we observed no obvious anomalies near $\gamma_n^{crit}$.  We did find that the density of the crystal increases with increasing dissipation, and that close $\gamma_n^{crit}$ the crystal is nearly close-packed.  

Finally, we investigated whether the melting transition is described by the Lindemann ratio \cite{lindemann}, the ratio of the rms fluctuation in particle positions in the solid phase to the lattice spacing.   Equilibrium solids typically melt when this ratio exceeds about 0.15, and the melting of a granular monolayer was found be be consistent with this criterion  \cite{reis06}.  We find that the Lindemann ratio at the melting line (the open squares in Fig. \ref{Gg}a) ranges from about 0.3 at  very low inelasticity ($\gamma_n=0.1$) to 0.2 at $\gamma_n=350$.   Due to the small size of the crystal close to the melting transition it is difficult to unambiguously measure the amplitude of the positional fluctuations, so these preliminary results will need to be confirmed with larger scale simulations.  We have observed previously that the diffusion coefficient in the gas phase is larger for more inelastic particles \cite{JCP}, presumably as a result of increased correlations.  The rms fluctuations in the solid phase also increases with increasing inelasticity for fixed acceleration (data not shown). However, the amplitude of the fluctuations decrease upon decreasing acceleration, and along the melting line the decrease from the acceleration is greater than the increase from the inelasticity, resulting in a reduction of the Lindemann ratio.  These results suggest that the effects of forcing and/or dissipation on the ordering transition result in a complexity that is not present in equilibrium systems, where the Lindemann ratio usually provides a reliable guide to the transition point.

These results  demonstrate that the phase behavior of steady states of inelastic spheres is considerably different than their equilibrium analogs.  The extension of the ideas of equilibrium statistical mechanics to nonequilbrium systems remains an ongoing challenge, and the insights gained from careful comparisons of closely analogous model systems are necessary to develop and test new approaches.  Our observation that ordering is suppressed by inelasticity for identical inelastic spheres shows that quantitative changes in the strength of dissipation can qualitatively alter the phase diagram of even the simplest granular media. 

\begin{acknowledgments}
This work was supported by NASA under award number NNC04GA63G. One of the authors (F. V. R.) also acknowledges financial support from the Spanish Ministry of Education and Science through "Juan de la Cierva" research program.
\end{acknowledgments}

\vspace{3 mm}


\end{document}